\documentclass[twocolumn,tighten]{aastex6}
\pdfoutput=1 
\usepackage{amsmath,amstext}
\usepackage[T1]{fontenc}
\usepackage{apjfonts} 
\usepackage[figure,figure*]{hypcap}

\shortauthors{Gasque et al.}
\shorttitle{Two Long-Period Cataclysmic Variables}

\begin{document}

\title{Two Long-Period Cataclysmic Variable Stars: ASASSN-14\MakeLowercase{ho} and V1062 C\MakeLowercase{yg}}

\author{L. Claire Gasque$^{1}$, Callum A. Hening$^{1}$, Raphael E. Hviding$^{1,2}$, John R. Thorstensen$^{1}$, Kerry Paterson$^{3}$, Hannes Breytenbach$^{3,4}$, Mokhine Motsoaledi$^{3,4}$, Patrick A. Woudt$^{3}$}

\affil{$^{1}$Department of Physics and Astronomy, Dartmouth College, Hanover, NH 03755, U.S.A.\\
$^{2}$Steward Observatory, University of Arizona, Tucson, AZ 85721, U.S.A.\\
$^{3}$Department Astronomy, University of Cape Town, Private Bag X3, Rondebosch 7701, South Africa \\
$^{4}$ South African Astronomical Observatory, PO Box 9, Observatory 7935, Cape Town, South Africa}

\begin{abstract} We report spectroscopy and photometry of the cataclysmic
variable stars ASASSN-14ho and V1062 Cyg.  Both are dwarf novae with spectra
dominated by their secondary stars, which we classify approximately as K4 and M0.5,
respectively.  Their orbital periods, determined mostly from the secondary
stars' radial velocities, proved to be nearly identical, respectively 
$350.14\pm0.15$ and $348.25\pm0.60$\,min.
The H$\alpha$ emission line in V1062 Cyg displays a relatively sharp emission
component that tracks the secondary's motion, which may arise
on the irradiated face of the secondary; this is not often seen and may 
indicate an unusually strong flux of ionizing radiation.  Both systems exhibit 
double-peaked orbital modulation consistent with ellipsoidal variation from the 
changing aspect of the secondary.  We model these variations to constrain the 
orbital inclination $i$, 
and estimate approximate component masses based on $i$ and the secondary
velocity amplitude $K_2$.
\end{abstract} 

\section{Introduction}
\label{sec:intro}
Cataclysmic variable stars (CVs) are binary stars in which a white dwarf primary accretes matter from a more 
extended secondary star via Roche lobe overflow.  The accreting matter usually forms a disk around the primary. 
A good overview of CVs is presented in \citet{Warner95CVs}.

Dwarf novae (DN), one of the most common types of CV, are weakly or
non-magnetic systems which can spontaneously go from quiescence into outburst,
a more luminous state corresponding to a brightening of several magnitudes.
DN outbursts are thought to be caused by the rapid release of gravitational potential energy
triggered by instabilities in the accretion disk.  At longer orbital
periods, the quiescent spectra of DN are often
dominated by the spectrum of a late-type secondary 
\citep{Warner95CVs}.    

DN systems in which the secondary star is the dominant flux contributor often
show a modulation in the light curve caused by the tidal distortions in the
secondary star \citep{Barwig83DwarfNovae}.  The projection effects
cause two maxima per orbital period at the quadrature phases, and gravity
darkening causes the side of the secondary facing the white dwarf to appear
darker than the back side, resulting in unequal minima
\citep{Bochkarev79ellipsoidality}. 

This work focuses on two DN systems: ASASSN-14ho and V1062 Cyg. We obtained
optical spectroscopy and photometry of these systems in order to determine 
their orbital periods and estimate the spectral
types of their secondary stars and the binary system parameters.

ASSASN-14ho (RA = 06$^\text{h}$30$^\text{m}$27.3$^\text{s}$, Dec =
$-65^\circ$29$''$50.2$'$, J2000) was announced as a CV in outburst 
following observations taken of the system on the du Pont 2.5\,m spectrograph
at the Las Campanas Observatory in September 2014 \citep{Prieto14ASN14ho}. The
system had quiescent brightness of $V = 15.4$ on 2014 Sept.~9 and reached
$V = 10.92$ on 2014 Sept.~11 Sept before declining.  A previous outburst
had been seen in 2009 by the Catalina Real Time Transient Survey 
\citep{Drake09CRTS}.  Inverting the Gaia Data Release 2 (DR2)
parallax \citep{Prusti16Gaia,GaiaDR2}
gives a distance estimate of $331 \pm 3$ pc.  

V1062 Cyg (RA = $21^\textrm{h} 08^\textrm{m} 12.20^\textrm{s}$, Dec =
$+36^\circ 49'' 27.7'$, J2000) was first reported by \citet{Hoffmeister1965New},
with a magnitude range of 15.5 - 18 mag.
Despite having since appeared in several CV catalogs, the system has apparently
never been studied in detail and its orbital period remained unknown.  The
inverse of its Gaia DR2 parallax is $840\,(+95,-78)$ pc.

We describe our observations, instruments, and analysis in Section \ref{sec:meth}. 
In Sections \ref{sec:14ho} and \ref{sec:62cyg} we present our results for ASSASN-14ho 
and V1062 Cyg, respectively. Finally, our conclusions and discussion are presented 
in Section \ref{sec:disc}.

\begin{deluxetable}{lrrrr}
\tablecolumns{3}
\tablewidth{2.0\columnwidth}
\tablecaption{\label{tab:14spec}Radial Velocities of ASASSN-14ho}
\tablehead{
\colhead{Time\tablenotemark{a}} & 
\colhead{$v_{\rm abs}$} &
\colhead{$v_{\rm emn}$} \\
\colhead{ } & 
\colhead{(km s$^{-1}$)} &
\colhead{(km s$^{-1}$)\tablenotemark{b}} \\
}
\startdata
57794.3628  & $  318 \pm   10$ & $  -45 \pm   13$ \\
57794.3698  & $  319 \pm   11$ & $  -35 \pm   14$ \\
57794.3771  & $  269 \pm   12$ & $  -55 \pm   16$ \\
57794.3841  & $  268 \pm   10$ & $  -39 \pm   15$ \\
57794.3984  & $  201 \pm   11$ & $  -62 \pm   14$ \\
57794.4054  & $  154 \pm   11$ & $  -43 \pm   16$ \\
57794.4124  & $  128 \pm   15$ & $   -6 \pm   16$ \\
57794.4193  & $   63 \pm   13$ & $   -7 \pm   21$ \\
57794.4287  & $   41 \pm   30$ & $   56 \pm   20$ \\
57794.4357  & $  -38 \pm   15$ & $   80 \pm   22$ \\
57794.4427  & $  -45 \pm   12$ & $   72 \pm   18$ \\
57794.4497  & $  -98 \pm   12$ & $   63 \pm   23$ \\
57794.4594  & $ -131 \pm   13$ & $   11 \pm   26$ \\
57794.4664  & $ -143 \pm   11$ & $   66 \pm   21$ \\
57794.4734  & $ -152 \pm   12$ & $   90 \pm   23$ \\
57794.4804  & $ -173 \pm   10$ & $   76 \pm   27$ \\
57794.4895  & $ -148 \pm   13$ & $  123 \pm   20$ \\
57794.4964  & $ -145 \pm   12$ & $  119 \pm   21$ \\
57794.5034  & $ -140 \pm   11$ & $   70 \pm   18$ \\
57794.5104  & $ -106 \pm   28$ & $  108 \pm   16$ \\
57794.5205  & $  -67 \pm   11$ & $  101 \pm   17$ \\
57794.5275  & $  -20 \pm   19$ & $  100 \pm   19$ \\
57794.5345  & $   18 \pm   12$ & $   57 \pm   25$ \\
57794.5415  & $   75 \pm   14$ &  \nodata  \\
57794.5498  & $  125 \pm   13$ &  \nodata  \\
57794.5572  & $  112 \pm   34$ & $   62 \pm   26$ \\
57794.5642  & $  208 \pm   15$ &  \nodata  \\
57794.5712  & $  237 \pm   19$ &  \nodata  \\
57796.2883  & $  300 \pm   11$ & $  -34 \pm   16$ \\
57796.2953  & $  311 \pm    9$ & $  -26 \pm   16$ \\
57796.3023  & $  302 \pm    9$ & $  -78 \pm   16$ \\
57796.3093  & $  316 \pm   10$ & $  -46 \pm   17$ \\
57796.3173  & $  290 \pm    9$ & $  -43 \pm   21$ \\
57796.3243  & $  271 \pm   10$ & $  -24 \pm   19$ \\
57796.3313  & $  246 \pm   10$ & $  -40 \pm   20$ \\
57796.3383  & $  218 \pm   10$ & $  -72 \pm   21$ \\
57796.3500  & $  164 \pm   11$ & $  -59 \pm   24$ \\
57796.3570  & $  104 \pm    9$ & $  -37 \pm   21$ \\
57797.2745  & $  295 \pm   10$ & $  -31 \pm   20$ \\
57797.2862  & $  267 \pm   10$ & $  -33 \pm   17$ \\
57797.2978  & $  254 \pm    8$ & $  -36 \pm   18$ \\
57799.2718  & $  150 \pm    8$ &  \nodata  \\
57799.2802  & $   94 \pm    8$ &  \nodata  \\
\enddata
\tablenotetext{a}{Barycentric JD of mid-integration, minus 2,400,000, in the UTC
time system.}
\tablenotetext{b}{Computed using a convolution function with positive and
negative gaussians separated by 44 \AA, each with full-width at half-maximum
(FWHM) of 7 \AA .} 
\end{deluxetable}
\begin{deluxetable}{lrrrrrr}
\tablecolumns{4}
\tablewidth{500pt}
\tablecaption{\label{tab:62spec}Radial Velocities of V1062 Cyg}
\tablehead{
\colhead{Time\tablenotemark{a}} & 
\colhead{$v_{\rm abs}$} &
\colhead{H$\alpha$ wing}\tablenotemark{b} & 
\colhead{H$\alpha$ peak}\tablenotemark{c} \\
\colhead{} \\
\colhead{(km s$^{-1}$)} &
\colhead{(km s$^{-1}$)} &
\colhead{(km s$^{-1}$)} &
}
\startdata
57928.7553  & $  139 \pm  48$ &  \nodata  & $112 \pm 34 $\\
57928.7640  & $   83 \pm  33$ &  \nodata  & $ 92 \pm 23 $\\
57928.7727  & $  171 \pm  30$ &  \nodata  & $ 49 \pm 22 $\\
57928.7814  & $  158 \pm  32$ &  \nodata  & $ 94 \pm 20 $\\
57928.7931  & $  110 \pm  31$ & $ -185 \pm 39$   &  $57  \pm 22$ \\
57928.8074  & $  146 \pm  26$ & $ -202 \pm 27$   &  $81  \pm 15$ \\
57928.8602  & $ -107 \pm  30$ & $  -45 \pm 31$   &  $-60 \pm 23$ \\
57928.8745  & $ -163 \pm  29$ & $   12 \pm 32$   & $-76 \pm 22$  \\
57928.8888  & $ -207 \pm  25$ & $  -36 \pm 25$  & $-147 \pm 16$ \\ 
57928.9031  & $ -227 \pm  23$ & $   -7 \pm 30$  & $-157 \pm 17$ \\ 
57928.9174  & $ -213 \pm  23$ & $   18 \pm 31$  & $-153 \pm 18$ \\ 
57928.9316  & $ -204 \pm  29$ & $   72 \pm 36$  & $ -66 \pm 20$ \\ 
57928.9459  & $  -99 \pm  18$ & $    3 \pm 28$  & $ -85 \pm 15$ \\ 
57928.9602  & $  -33 \pm  25$ & $  -43 \pm 33$  & $ -90 \pm 21$ \\ 
57929.8507  & $ -160 \pm  28$ & $   34 \pm 38$  & $-221 \pm 19$ \\ 
57929.8684  & $ -215 \pm  22$ & $   81 \pm 24$  & $-143 \pm 14$ \\ 
57929.8862  &  \nodata  &  \nodata   & $ -80 \pm  31$ \\
57930.7428  &  \nodata  &  $ -147 \pm 30$ & \nodata  \\ 
57930.7605  & $   72 \pm  35$ & $ -186 \pm  25$  & $31 \pm 30$ \\ 
57930.7783  & $  -81 \pm  44$ &  \nodata  & \nodata  \\ 
57930.7960  & $ -153 \pm  46$ & $  -75 \pm  34$  & $-172 \pm 33$ \\ 
57930.8137  & $ -189 \pm  29$ & $  -41 \pm  30$  & $-224 \pm 20$ \\
57930.8331  & $ -196 \pm  33$ & $   67 \pm  25$  & $-184 \pm 21$ \\ 
57930.9336  & $   93 \pm  27$ & $ -198 \pm  19$  & $68  \pm 23$ \\ 
\enddata
\tablenotetext{a}{Barycentric JD of mid-integration, minus 2,400,000, in the UTC
time system.}
\tablenotetext{b}{Radial velocity of the H$\alpha$ line wings, measured using a 
convolution function consisting of positive and negative gaussians with FWHM
10 \AA , separated by 32 \AA .}
\tablenotetext{c}{Radial velocity of the H$\alpha$ line peak, measured using
a convolution function formed from the derivative of a gaussian, optimized for a
line with FWHM 12 \AA .}
\end{deluxetable}

\begin{table}[ht]
\centering
{\begin{center} \caption{Photometric Observations of ASASSN-14ho \label{tab:14phot}}
\begin{tabular}{ccccccc}
\hline
\hline
Date & Time & Exp & N & \multicolumn{2}{c}{Airmass}\\
&(UTC)&(s)&&Start&End&\\
\hline
2017-02-11&18:44:21&10.0&450&1.2079&1.1954\\
2017-02-11&19:59:41&15.0&320&1.1954&1.2426\\
2017-02-11&21:20:10&25.0&355&1.2429&1.5361\\
\hline
\end{tabular}
\end{center}}
\end{table}

\begin{table}[ht]
\centering
{\begin{center} \caption{Photometric Observations of V1062 Cyg\label{tab:62phot}}
\begin{tabular}{ccccccc}
\hline
\hline
Date & Time & Exp & N & \multicolumn{2}{c}{Airmass}\\
&(UTC)&(s)&&Start&End&\\
\hline
2017-06-27&05:15:57&60.0&360&1.991&1.030\\
2017-06-28&05:03:12&60.0&360&2.094&1.023\\
\hline
\end{tabular}
\end{center}}
\end{table}

\section{Methods}
\label{sec:meth}

\subsection{Observations}
\label{ssec:datcoll}
Our observations of ASASSN-14ho were taken in February 2017  at the
South African Astronomical Observatory (SAAO) near Sutherland.  Observations of V1062 Cyg were taken in June 2017 at MDM Observatory, on Kitt Peak, Arizona.



Our spectra of ASASSN-14ho were taken with the Spectrograph Upgrade -- Newly
Improved Cassegrain (SpUpNIC) instrument \citep{Crause16Spupnic} mounted on the
SAAO 1.9\,m telescope. We used the G6 grating (blazed for 4600\,\AA ) which
gave a dispersion of 1.36\,\AA\,px$^{-1}$. A 1.35-arcsec slit and a grating
angle of 11.75$^\circ$ yielded a resolving power of 1000 from 4200 to
7000\,\AA. We obtained 43 spectra of ASASSN-14ho, mostly 600-s exposures, over
five days, for a total integration time of just over 7.5 hr. Table
\ref{tab:14spec} gives the times of observation and radial velocities derived
from the spectra (described in Section \ref{ssec:anal}).

The V1062 Cyg spectra were taken with the
modspec\footnote{http://mdm.kpno.noao.edu/Manuals/ModSpec/modspec\_man.html}
spectrograph on the 2.4\,m Hiltner telescope. We used a 600\,mm$^{-1}$ 5000
\AA\ blazed grating and a $2048\times 2048$ 24-micron pixel SITe CCD, which
gave a wavelength range of 4300 \AA \ to 7500 \AA \ and a resolution of
3.4\,\AA . Table \ref{tab:62spec} gives the times of observation and radial
velocities derived from the spectra.

The photometry of 
ASASSN-14ho is from the Sutherland High Speed Optical Camera 
(SHOC; \citealt{Coppejans13SHOC}) on the SAAO 1.0\,m telescope. The instrument
uses a back-illuminated frame transfer CCD with a 6.76\,ms dead time. The
system offers a $2.85'\times 2.85'$ field of view with a plate scale of
$0.668''\,\text{px}^{-1}$ in the $4\times4$ binning mode. Our observations, all from 2017 Feb. 11, consist of 1125 unfiltered frames obtained over five hours.  Table \ref{tab:14phot} gives the times of observations, length of exposure, number of frames taken, and airmass.

Our photometry of V1062 Cyg was taken with an Andor Ikon DU-937N
camera on the MDM 1.3\,m McGraw-Hill Telescope. The camera has $512\times512$
13\,$\mu\textrm{m}^2$ pixels which gave a
$2.32'\times2.32'$ field of view \citep{Thorstensen2013Andor}. Over the course
of two nights, 720 frames were taken for a total integration time of
twelve hours. Table \ref{tab:62phot} gives the times of observations, 
length of exposure, number of frames taken, and airmass. 

\subsection{Data Reduction}
\label{ssec:datred}


We reduced our spectra using procedures broadly similar to those outlined in
\citet{Thorstensen16BinaryZoo}.  We used
IRAF\footnote{IRAF is distributed by the National Optical Astronomy
Observatory, which is operated by the Association of Universities for Research
in Astronomy (AURA), Inc., under cooperative agreement with the National
Science Foundation.} for bias and flat field corrections.  For the 
SAAO spectra we extracted background-subtracted 1-dimensional spectra using
the IRAF {\it apall} task, and applied wavelength calibrations derived
from Cu-Ar arc spectra interleaved with the stellar exposures. 
For the MDM data, we extracted the spectra using a local implementation
of the algorithm given by \citet{horne86}, and a wavelength scale derived
by rigidly shifting a master pixel-to-wavelength solution using the
$\lambda 5577$ airglow line as a fiducial.
At both observatories we derived a flux calibration using observations
of standard stars taken during the run.


Both the SAAO SHOC and the MDM Andor camera write 3-dimensional
data cubes.  To process these we used python scripts to (1) create 
average bias images and flat field frames from exposures of the 
twilight sky (2) subtract the bias and divide by the flat field,
and (3) derive barycentric times of the exposure centers.  We then
split the individual frames from the data cube and 
measured the magnitudes of the program object
and several comparison stars using the aperture photometry 
task in the IRAF implementation of DAOPHOT \citep{Stetson8DAOPHOT},
and finally compiled the differential magnitudes into a time series.
The uncertainties in our program star measurements were estimated from 
the scatter in the differential magnitudes of apparently constant
stars of similar brightness. 

Our data cannot be transformed accurately to a standard passband
because the SAAO data were taken unfiltered, and the filter used
at MDM was intended only to mitigate scattered moonlight (passing
$\lambda > 4200$\ \AA).  However, we did transform to approximate $V$ magnitude
by adding the $V$ magnitudes of our comparison stars to the differential
magnitudes.  These pseudo-$V$ magnitudes should not be very far
off, since in both systems the late-type secondary star dominates
the light; the color mismatch between the CV and the comparison star is
therefore unlikely to be severe.

\subsection{Analysis}

\label{ssec:anal}

We measured radial velocities of the secondary stars using the cross-correlation technique of 
\citet{Tonry79redshifts}. For 
the template spectrum, we used the average of 76 spectra of G and K type stars, taken
with the Hiltner telescope and modspec in the same configuration used here for V1062 Cyg,
that had been shifted to zero apparent velocity prior to averaging. 

To measure emission-line velocities, we used a convolution technique developed
by \citet{Schneider80Emis}, in which the line profile is convolved with an odd
function and the zero of the convolution is taken as a measure of the line
center.  For the odd function, we used either the derivative of a gaussian 
(`dgau' for short) or the sum of positive and negative
Gaussians, offset by a selectable width (`gau2').  In both cases, only the H$\alpha$
line gave usable velocities.  


To search for periods, we fit general least-squares sinusoids to the radial
velocities at each frequency in a dense grid of trial frequencies.  In both
stars studied here, the adopted orbital frequency had by far the lowest
scatter aound the best fit, and there was no ambiguity in
the daily cycle count.  We fit the radial velocities with functions of the form
\begin{equation} \nonumber v(t) = K\sin\left(\frac{2\pi(t-T_0)}{P}\right) +
\gamma \end{equation} \noindent where $P$ is the orbital period and $T_0$ is an
epoch of blue-to-red crossing chosen to be near the center of the time series.
If the velocities faithfully trace the motion of one component, then $\gamma$
will be the systemic radial velocity and $K$ will be the line-of-sight
component of that star's orbital velocity.  We used a Monte Carlo simulation to
estimate the uncertainties in the parameters of our best-fitting sinusoid.

To estimate the secondary stars' spectral types, we began by shifting our
individual flux-calibrated spectra to the secondary's rest frame
and averaging them.
We then subtracted spectra of known spectral type taken from the same instrument, varying
the spectral type and normalization with the aim of retrieving
a smooth continuum as judged by eye. 
This resulted in a range of possible classifications and V-band
magnitudes for the secondary. 

To display the spectrum as a function of orbital phase, we set up a grid of 100
evenly-spaced fiducial phases around the orbital cycle, and computed average
spectra for each phase.  Weights for the spectra near each fiducial phase were
computed using a Gaussian weighting function, with a standard deviation of
$0.03$ cycles.  The spectra were rectified (divided by a continuum fit) and
cleaned of obvious artifacts before being averaged.  We then stacked the
phase-averages into a 2-dimensional image, and generated greyscale figures from
these.


The light curves of both objects showed two peaks per
orbit and unequal minima, as expected from the changing
aspect of a tidally distorted secondary star. To explore
this, we used a modeling program described by \cite{ta05}.
The program models the Roche-lobe-filling 
secondary star as a 500-sided
polyhedron and computes the $V$-band light curve 
for the source as observed from Earth. The model parameters 
are the (assumed known) orbital period, the masses of both stars, 
the orbital inclination $i$, the secondary's effective temperature $T_{\rm eff}$, 
limb- and gravity-darkening coefficients, the distance and 
reddening of the system, and an assumed flat (in flux per
unit wavelength $F_{\lambda}$) 
`extra light' contribution from the primary and accretion structures.
Our estimates of $T_{\rm eff}$ and the `extra light' are guided by the spectral
decompositions used to estimate the secondary's spectral type.
For all calculations we used a linear limb-darkening coefficient of 0.78,
from the tabulations of \citet{Claret2012} for $T_{\rm eff} = 4500$ K and
the Kepler passband, which is similar to our observed bands.  

From $P$ and the secondary star's velocity semiamplitude 
$K_2$, we compute the mass function
\begin{align*}
    f = \frac{(M_1)^3}{(M_1+M_2)^2}\sin^3({i}) = \frac{P}{2\pi G}(K_2)^3.
\end{align*}
We use this to constrain the secondary star's mass $M_2$ by estimating a range
of plausible inclinations $i$ from the light curve model fits, and assuming
that the white dwarf mass $M_1$ lies in a reasonable range
\citep{Kepler07WhiteDwarfMasses,Ritter86DwarfMass,Schreiber2016primaryM}.
Sections \ref{ssec:amp} and \ref{ssec:phot62cyg} detail this process for for
ASASSN-14ho and V1062 Cyg respectively.

\section{ASASSN-14\MakeLowercase{ho} Results}
\label{sec:14ho}

\subsection{Period and Radial Velocities} \label{ssec:14hoper} 
\begin{figure}[ht]
\centering
\includegraphics[trim=1cm 0 0 0 0,scale = .39]{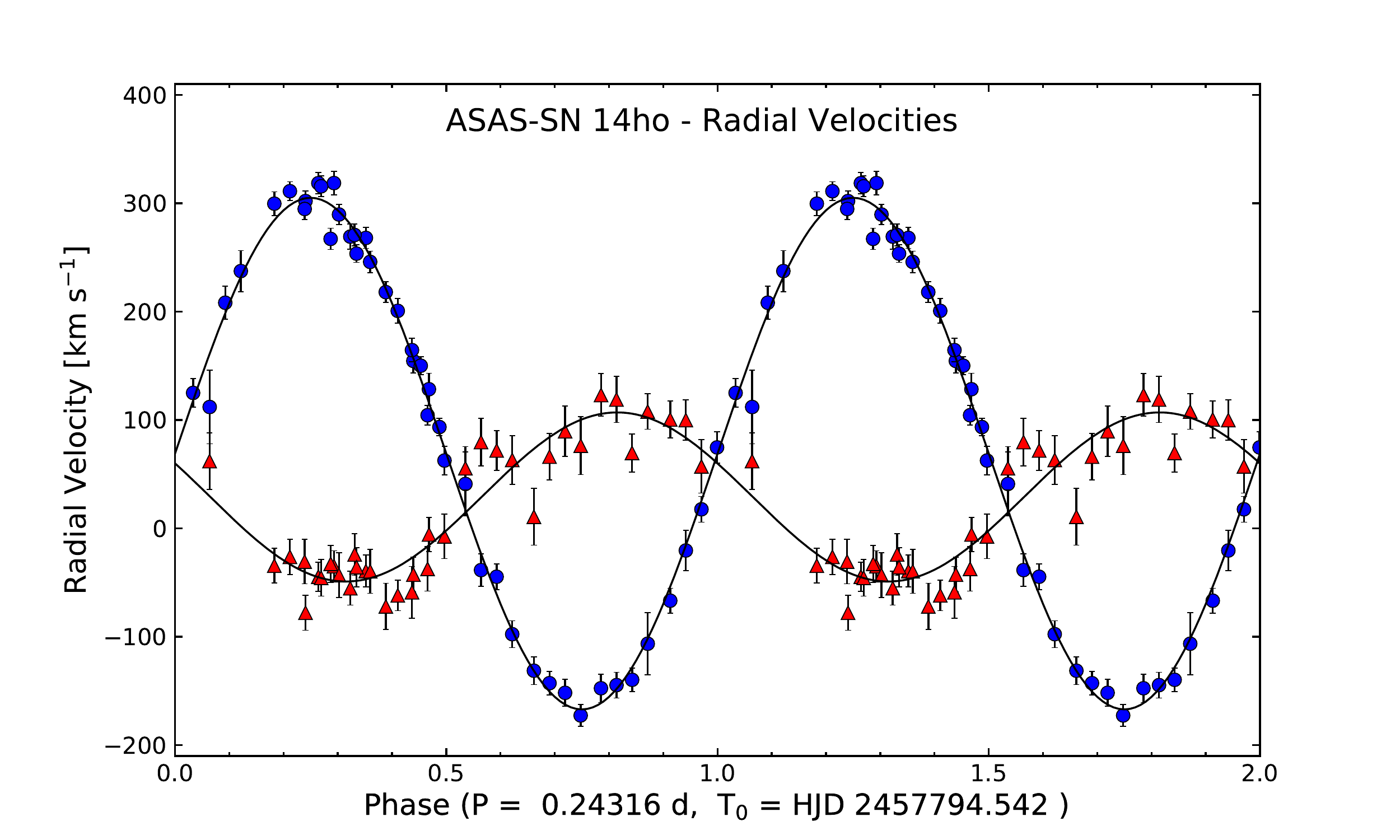}
\caption{\small ASASSN-14ho radial velocities folded on the orbital period, with one cycle
repeated once for clarity.  The blue circles are absorption-line velocties of the secondary
star, and the red triangles are velocities of the H$\alpha$ emission line wings.}
\label{fig:spec14ho}
\end{figure}

The absorption
line radial velocities are modulated with an unambiguous period of
$350.14 \,\pm\,0.15$\,min, which we adopt as $P_{\rm orb}$.  The H$\alpha$ 
line is double-peaked with a full width at half-maximum of $\sim 31$ \AA.
We measured its radial velocity using a `gau2' convolution function 
with a separation of 44 \AA, which emphasized the steep sides and
wings of the line.  The emission velocities showed modulation at 
essentially the same period as the absorption, but with greater
scatter.
Table \ref{tab:parameters} gives parameters of sinusoidal fits to the emission
and absorption velocities, and Figure \ref{fig:spec14ho} shows the 
velocities plotted as a function of phase, along with the best-fit
sinusoids. 


\begin{deluxetable*}{lllrrcc}
\tablecolumns{7}
\footnotesize
\tablewidth{0pt}
\tablecaption{\label{tab:parameters} Fits to Radial Velocities}
\tablehead{
\colhead{Data set} & 
\colhead{$T_0$\tablenotemark{a}} & 
\colhead{$P$} &
\colhead{$K$} & 
\colhead{$\gamma$} & 
\colhead{$N$} &
\colhead{RMS}  \\ 
\colhead{} & 
\colhead{} &
\colhead{(d)} & 
\colhead{(km s$^{-1}$)} &
\colhead{(km s$^{-1}$)} & 
\colhead{} &
\colhead{(km s$^{-1}$)}
}
\startdata
ASASSN-14ho absorption & 2457794.5417(10)  & 0.24316(10) & 236(5) & 69(4) & 43 & 13 \\
ASASSN-14ho emission   & 2457794.436(7)  & [0.24316]\tablenotemark{b}   &  78(10) & 29(8) & 37 & 22\\[1.5ex]
V1062 Cyg absorption   & 2457929.931(2)\tablenotemark{c} & 0.2418(5)\tablenotemark{c}   & 184(7) & $-$32(5) & 22 & 24 \\
V1062 Cyg H$\alpha$ cores & 57929.927(9) &  0.2418\tablenotemark{b} & 126(20) & $-37(17)$ & 23 &  41 \\ 
V1062 Cyg H$\alpha$ wings & 57929.818(7) &  0.2418\tablenotemark{b} &  129(19) & $-88(15)$ & 18 &  38 \\
\enddata
\tablenotetext{a}{Barycentric Julian date of mid-integration minus 2,440,000.  The time system is UTC.}
\tablenotetext{b}{Period fixed to the value derived from the absorption-line fit.}
\tablenotetext{c} {Light curves taken on subsequent nights suggest a slightly earlier $T_0$ and a slightly shorter $P$ than the best fit to the velocities (section \ref{ssec:phot62cyg}).}
\end{deluxetable*}

The systemic radial velocity
$\gamma$ derived from the absorption lines differs by 42 km s$^{-1}$ from the value from
the emission lines. We do not think this discrepancy is significant,
because the
emission line velocities do not
necessarily trace the white dwarf motion
\citep{North02velocities}.
If the emission originated directly from the white dwarf, we would expect an
emission-absorption phase offset of 0.5 cycle.  Our fitted sinusoids show a difference of
$0.56\pm0.03$ cycles, suggesting that
the emission from the accretion
disk does not track the white dwarf's position exactly.  The emission velocities
are uncertain enough that we are reluctant to use them to compute a mass ratio.

\begin{figure}[ht]
\centering
\includegraphics[scale = .5]{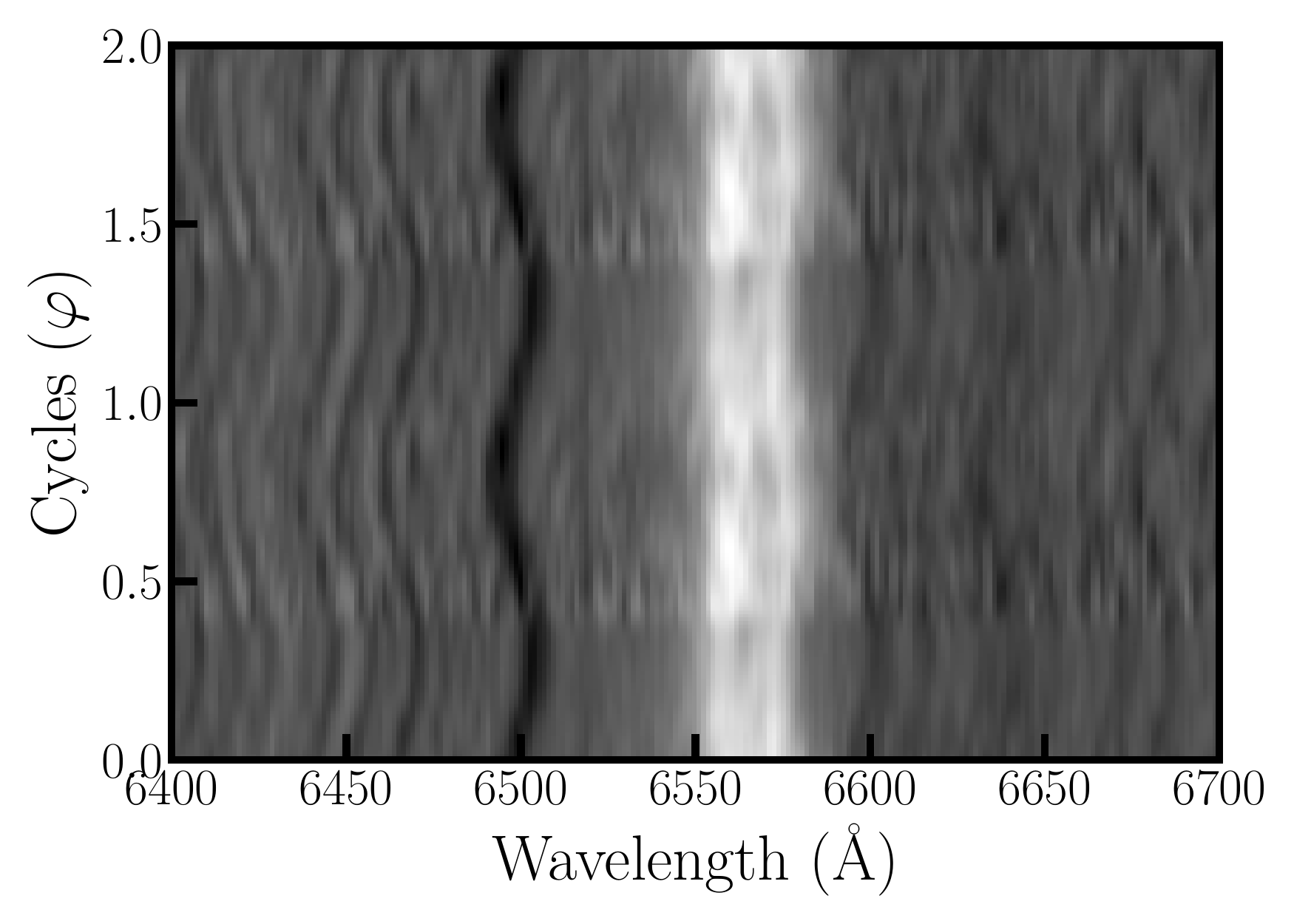}
\caption{\small Single-trailed spectrum on ASASSN-14ho, centered on the H$\alpha$ emission line.}
\label{fig:trail14ho}
\end{figure}

Figure \ref{fig:trail14ho} shows a single-trailed plot of the spectral features
of ASASSN-14ho, centered on the H$\alpha$ emission line. 
The H$\alpha$ profile is complex, but has the distinct double-peaked
structure characteristic of an accretion disk.  

\subsection{Spectral Type}
\label{ssec:14hospec}

\begin{figure}[ht!]
\centering
\includegraphics[scale = 0.47,trim=1.3cm 0 0 0]{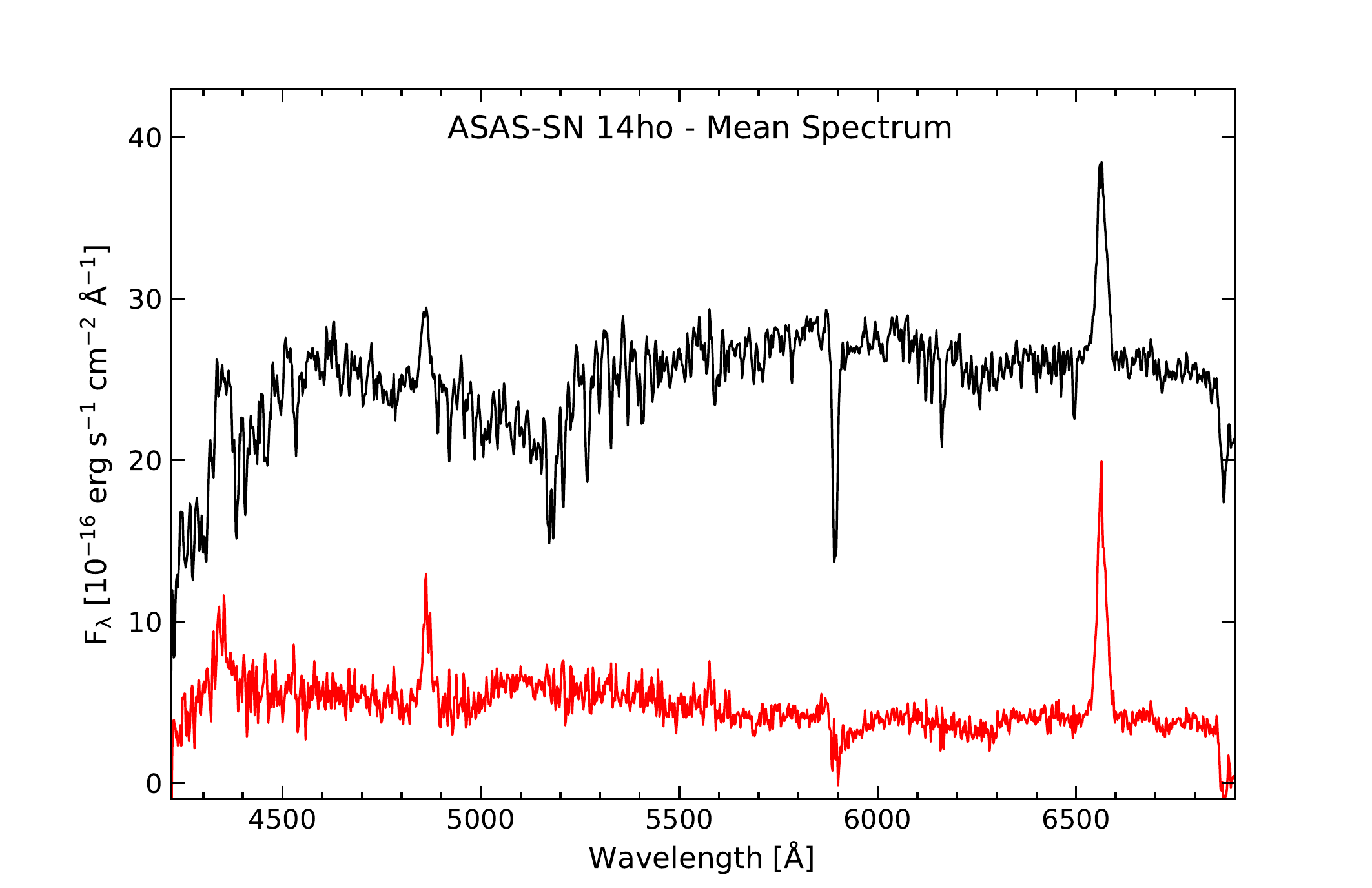}
\caption{\small Average spectrum of ASASSN-14ho (black) together with the
the spectrum after subtraction (red) of a scaled spectrum of the K4V star
Gliese 570a.  The individual spectra of ASASSN-14ho were shifted to the rest
frame of the secondary star before they were summed.}
\label{fig:asn14ho_primary}
\end{figure}

We find the secondary's spectral type to be near K4V, with an estimated uncertainty
of $\pm 2$ subclasses.  The secondary star contributes most of the light in the visible 
(Fig.~\ref{fig:asn14ho_primary}). The subtracted spectrum is characterized by a fairly
flat continuum with broad hydrogen emission lines, typical of dwarf novae at minimum light
\citep{Warner95CVs}.

\subsection{Photometry and Stellar Parameters}
\label{ssec:amp}

\begin{figure}[ht]
\centering
\includegraphics[trim=1cm 0 0 0,scale = 0.45]{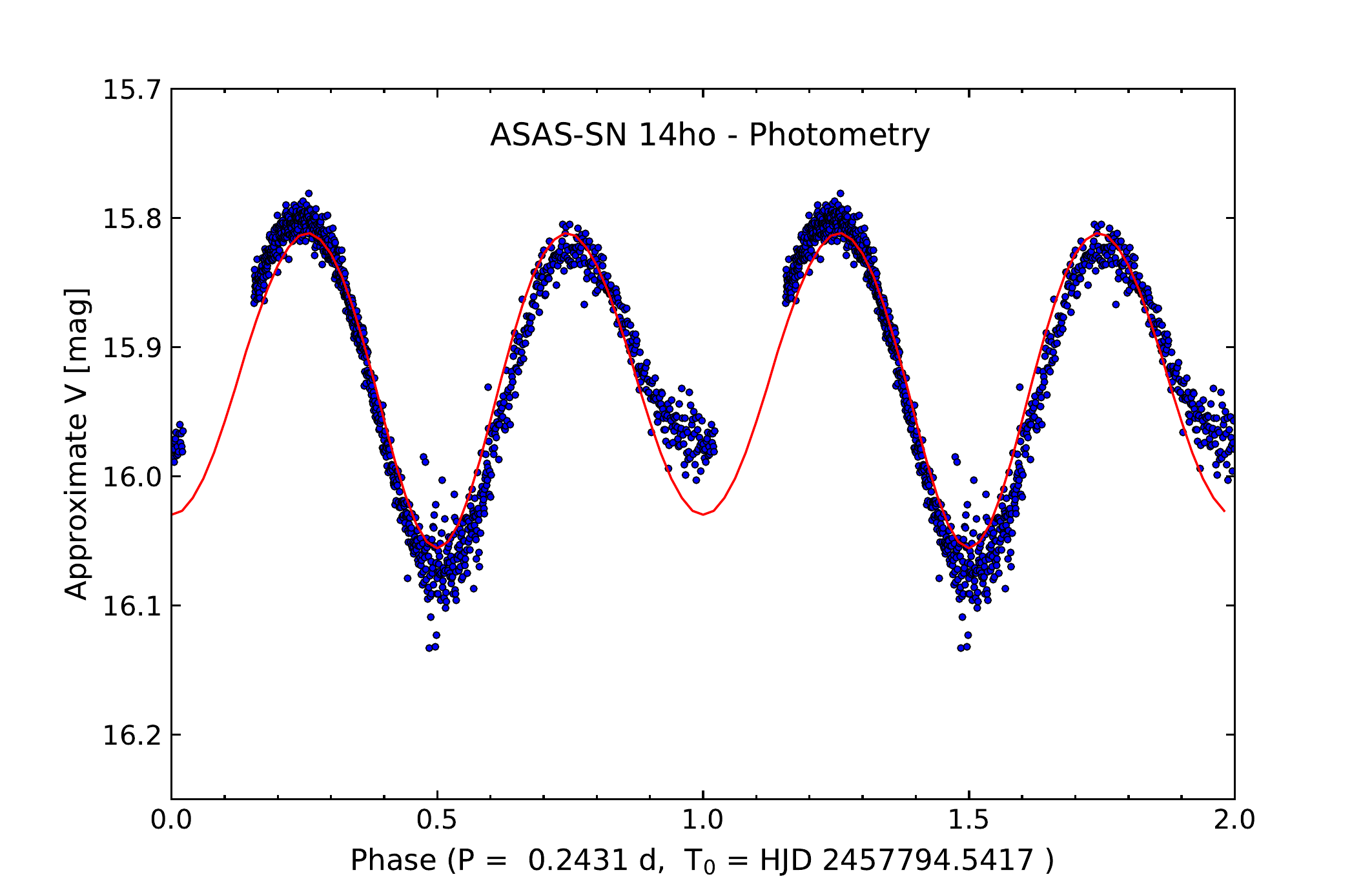}
\caption{\small Differential magnitudes of ASASSN-14ho transformed
to approximate $V$ plotted as a function of orbital phase,  
repeated once for clarity. The red curve shows a model of the ellipsoidal
variation.} 
\label{fig:asn14hophot}
\end{figure}

Figure~\ref{fig:asn14hophot} shows our photometry folded on the period and phase
determined in Section \ref{ssec:14hoper}.  As noted earlier, we transformed our instrumental magnitudes
to rough apparent $V$ magnitudes using $V = 15.25$ for the
comparison star 
\footnote{$\alpha$ = 6:30:24.87, $\delta$ = $-$65:31:21.3},
taken from the AAVSO Photometric All-Sky Survey (APASS) catalog 
\citep{henden15}.  
The light curve shows an obvious modulation with two peaks per orbit and unequal minima, as expected from a tidally-distorted secondary (so-called ellipsoidal variation), and no evidence of an eclipse.  
The red curve in Fig.~\ref{fig:asn14hophot} shows a light curve generated by 
the light-curve modeling program described earlier.  We could not fit all the
features in the light curve, 
such as the extra brightness near phase zero, with perfect fidelity.
Therefore, we did not attempt a formal best fit, but
adjusted the fit parameters by hand.  The fit shown has
$M_1 = 1.00$ and $M_2 = 0.28$ M${_\odot}$, $i = 52$ degrees, 
a secondary effective temperature of 4200 kelvin, and extra light
equivalent to $F_\lambda = 5 \times 10^{-16}$ erg s$^{-1}$ cm$^{-2}$ \AA$^{-1}$,
which gives a reddening-corrected distance nearly identical to the inverse
of the Gaia DR2 parallax, and matches the observed mass function. 
The amplitude of the modeled curve is influenced mostly by the inclination, the estimated contribution from the disk, constant ``extra light'' term, and the gravity darkening coefficient. With the last two factors held constant, we found reasonable fits in the range $i = 52 \pm 5$ degrees.  Varying $M_2$ mostly affects the normalization, because the Roche lobe radius scales roughly as 
$M_2^{1/3}$.

\begin{figure}[ht]
\centering
\includegraphics[scale = .37]{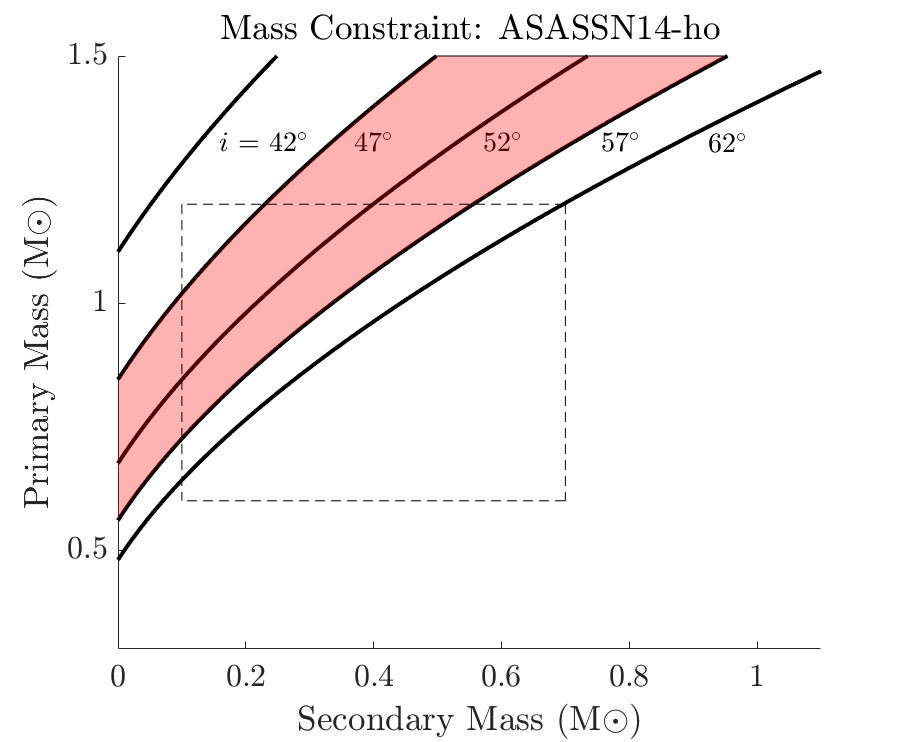}
\caption{\small Constraining masses of ASASSN-14ho using mass function. The shaded area represents the inclination range we report as most reasonable based on the photometry modeling. The dotted box represents a reasonable range of masses for each component in the system.}
\label{fig:14ho_mass}
\end{figure}

In Figure \ref{fig:14ho_mass}, we show the constraints on $M_1$ and $M_2$ set by the velocity amplitude,
for several inclination angles consistent with the light curve.  The
masses are sensitive to small changes in inclination. The box in the 
figure represents a rough range of reasonable masses for both components. 
If our inclination estimate is correct, the white dwarf mass is likely to be 
greater than $\sim 0.8$ solar masses and the secondary mass is likely to 
be rather low.

\section{V1062 Cyg Results}
\label{sec:62cyg}

\subsection{Period and Radial Velocities}
\label{ssec:62cygper}

\begin{figure}[ht]
\centering
\includegraphics[trim=1cm 0 0 0,scale = .39]{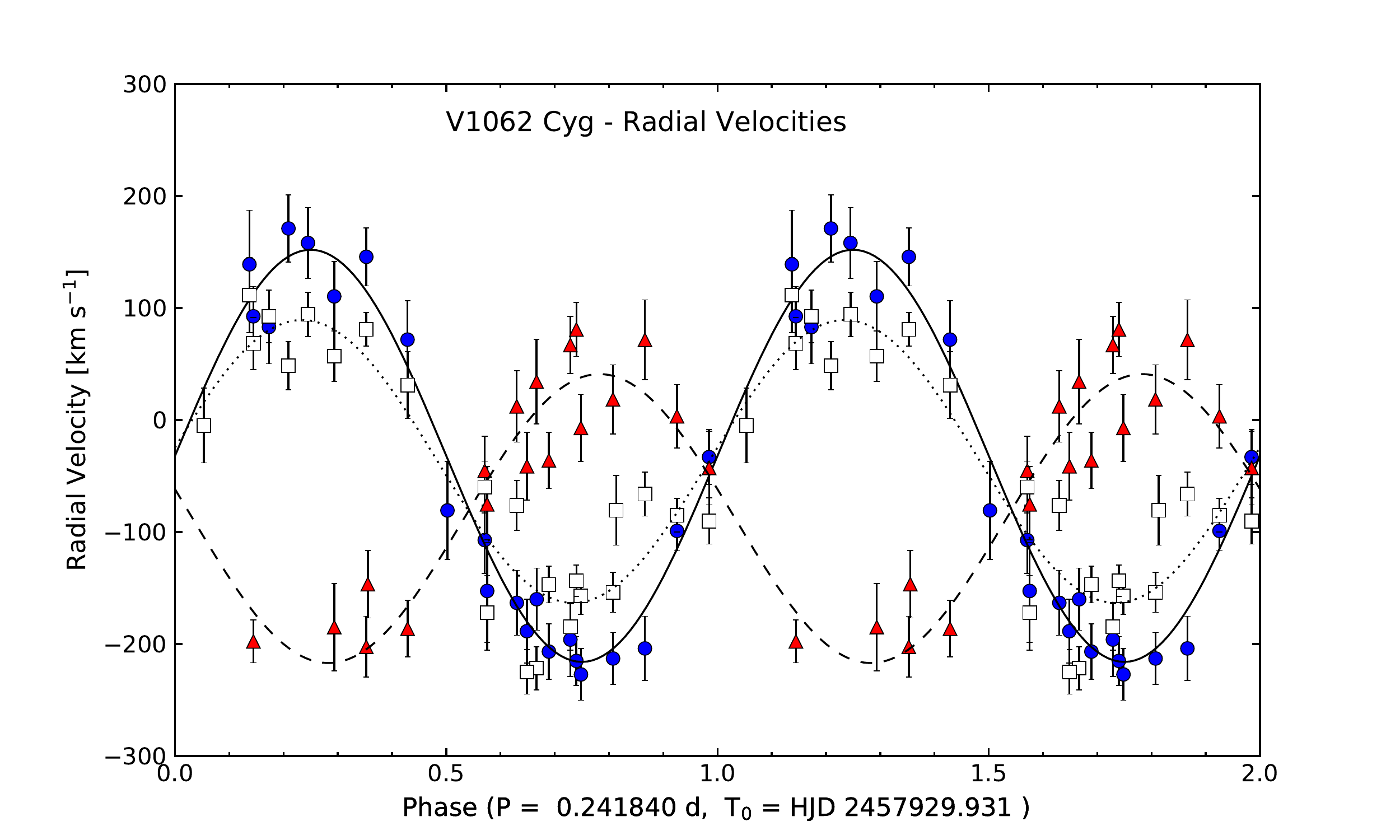}
\caption{\small Radial velocities from V1062 Cyg folded on the
adopted orbital period and repeated once for clarity, together with
the best-fitting sinusoids at the adopted orbital period.  The
symbols are as follows: 
Blue circles and solid line; absorption-line velocities of the 
secondary star.  Red triangles and dashed curve; H$\alpha$ emission
line wings; open squares and dotted curve; H$\alpha$ emission line
core.}
\label{fig:v1062cyg_foldedvs}
\end{figure}

The absorption line velocities in V1062 Cyg give an unambiguous $P_{\rm orb} =
348.25\,\pm\,0.84$\,min (note that the photometry discussed in
\ref{ssec:phot62cyg} suggests a slightly shorter period). 
The emission-line velocities corroborate the period found in the
absorption lines.  The H$\alpha$ emission line showed a complex structure, with 
the wings and core of the line moving roughly in antiphase.  To measure
the line wing velocities we used a `gau2' function separated by 32 \AA , 
and for the core we used a `dgau' optimized for 12 \AA\ FWHM.
Table \ref{tab:parameters} gives parameters of
least-squares sinusoidal fits, and Figure \ref{fig:v1062cyg_foldedvs} plots the
velocities of the absorption lines and the two components of the
H$\alpha$ line as a function of orbital phase, together with the best fits.

\begin{figure}[ht]
\centering
\includegraphics[scale = .5]{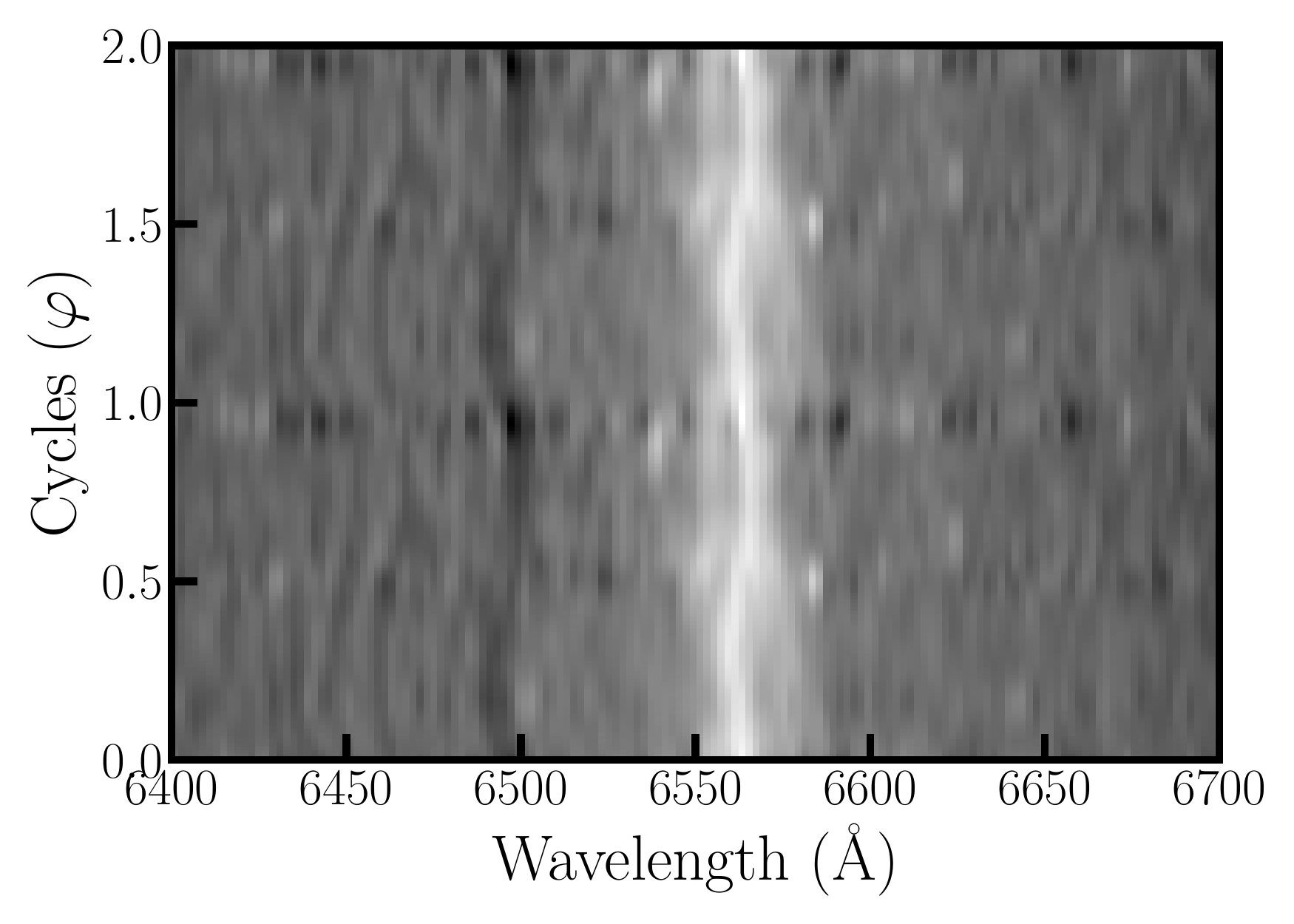}
\caption{\small Single-trailed spectrum on V1062 Cyg, centered on the H$\alpha$ emission line.}
\label{fig:v1062cyg_trailfig}
\end{figure}

Figure
\ref{fig:v1062cyg_trailfig} shows a phase-resolved greyscale representation of the spectrum centered on the H$\alpha$ emission line.
The velocity modulation of the bright line core is nearly in phase with the absorption, 
while the weaker, more diffuse wing component appears to move in antiphase.
This suggests that the core emission arises from the secondary star, caused perhaps from 
irradiation of the side facing the white dwarf  \citep{Steeghs01irradiation}, or possibly from magnetic
activity.  The velocity modulation of the wings is
apparently consistent with the accretion disk.

\subsection{Spectral Type}
\label{ssec:62cygspec}

\begin{figure}[ht]
\centering
\includegraphics[scale=0.47,trim=1.3cm 0 0 0]{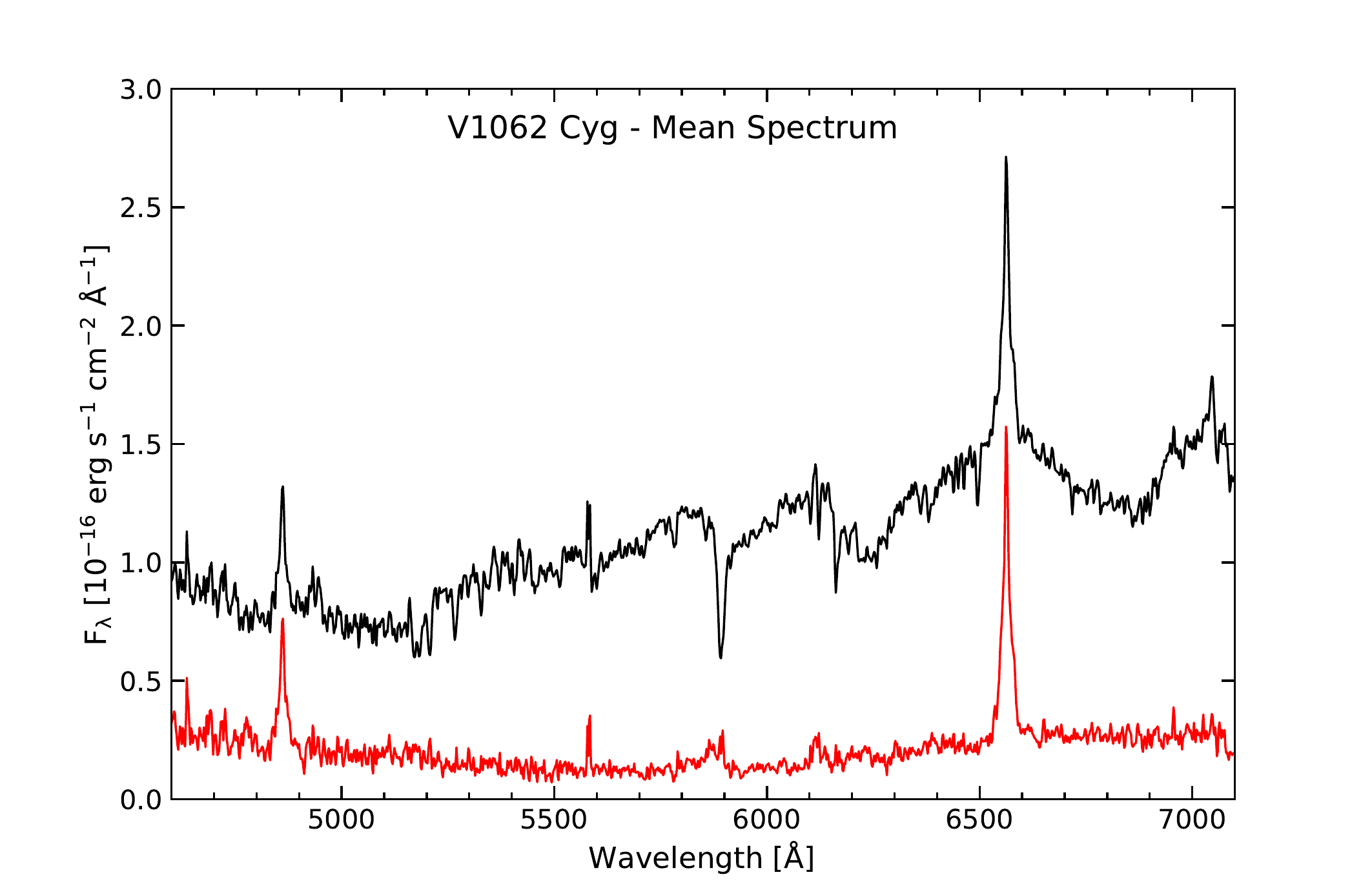}
\caption{\small Average spectrum of V1062 Cyg; the lower trace (red) shows the spectrum after subraction of a scaled spectrum of Gliese 486 (type M0.5V).}
\label{fig:v1062cyg_primary}
\end{figure}

The spectral type of the secondary is near M0.5. The total and subtracted spectra of V1062 Cyg are presented together in Figure \ref{fig:v1062cyg_primary}. 

\subsection{Photometry and Stellar Parameters}
\label{ssec:phot62cyg}

\begin{figure}[ht]
\centering
\includegraphics[trim=1cm 0 0 0,scale = 0.45]{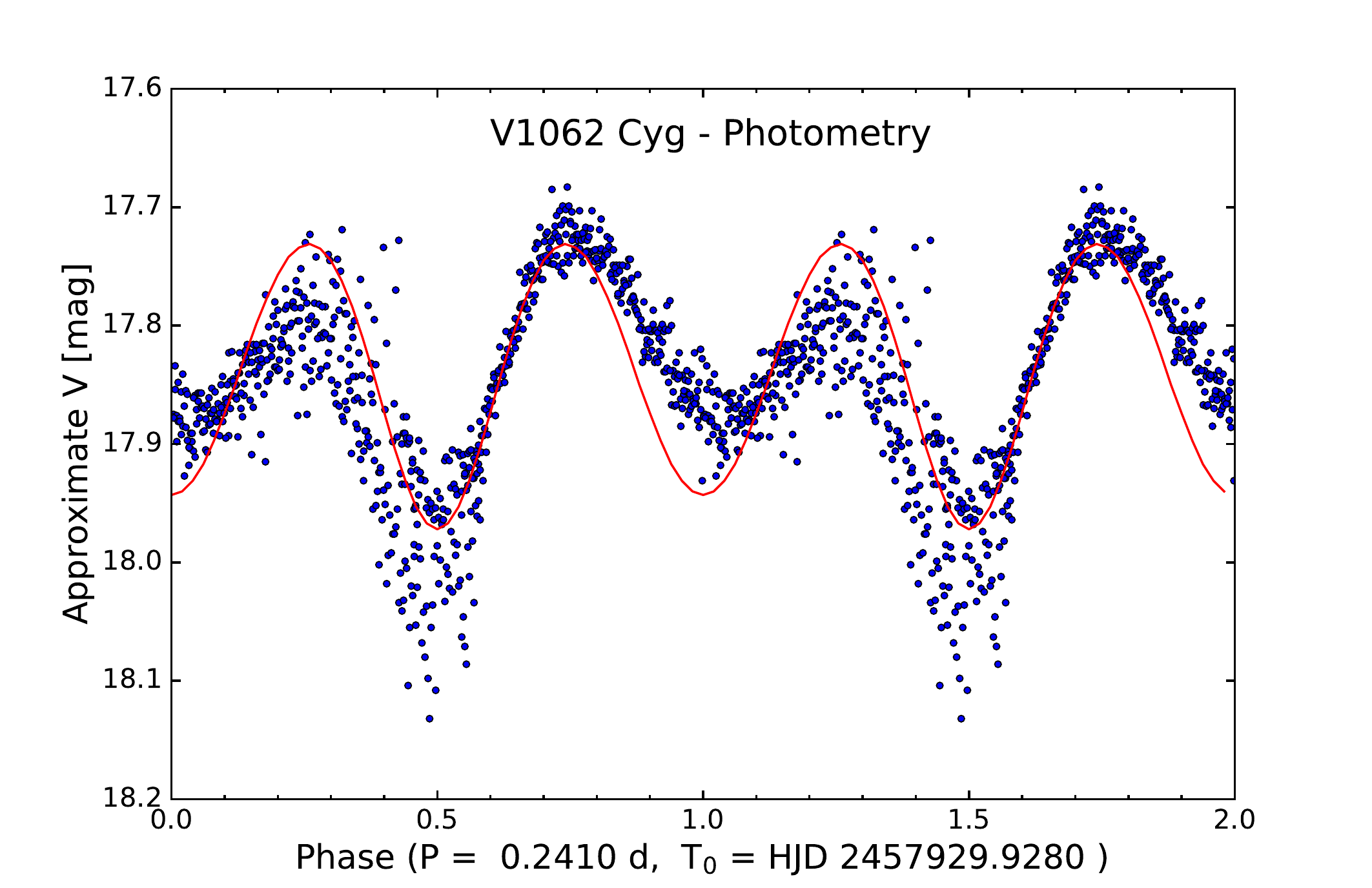}
\caption{\small Differential magnitudes of V1062 Cyg transformed
to approximate $V$ plotted as a function of orbital phase,  
repeated once for clarity. The red curve shows a model of the ellipsoidal
variation.} 
\label{fig:v1062cygphot}
\end{figure}

\begin{figure}[ht]
\centering
\includegraphics[scale = .37]{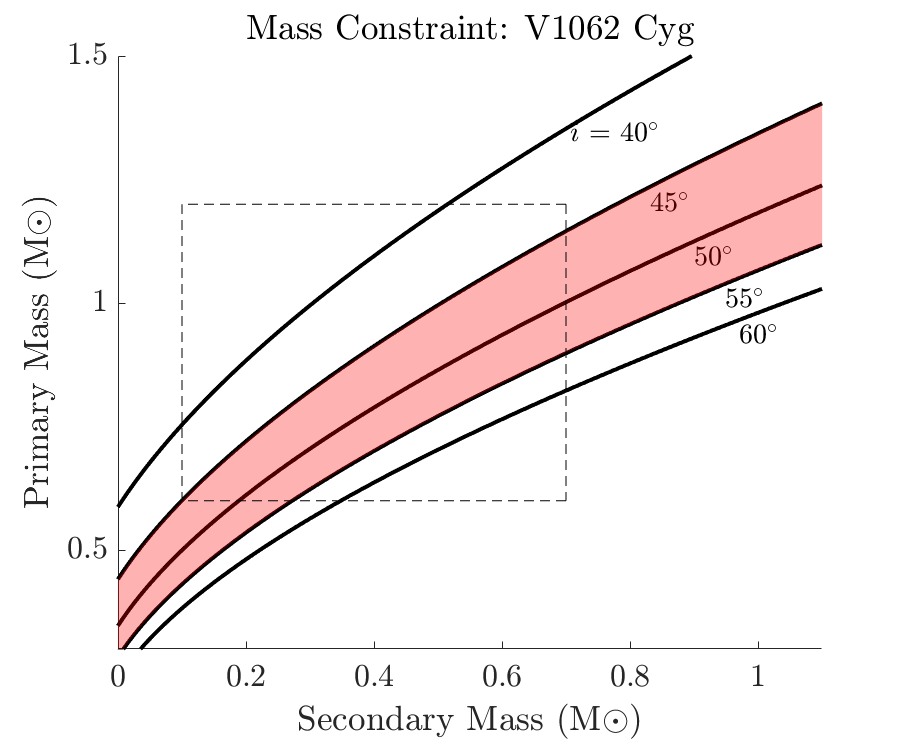}
\caption{\small Mass constraints for V1062 Cyg. The shaded area indicates the inclination range we estimate from the light-curve model. The dotted box represents the reasonable range of masses for each component in the system.}
\label{fig:62cyg_mass}
\end{figure}

Figure \ref{fig:v1062cygphot} shows the light curve, which displays double-peaked structure similar to
ASSASN-14ho (Section \ref{ssec:amp}).  The ellipsoidal model does not fit
as well as for ASASSN-14ho, but again clearly accounts for most of the variation\footnote{These data were taken several nights after the spectra; when we computed phases using the absorption-velocity ephemeris, the observed 
ellipsoidal humps arrived earlier than expected, by $\sim 0.04$ cycle.
Adjusting the period to a slightly shorter value, and the epoch $T_0$ to
be a bit earlier, corrected this mismatch; the values used are given in 
the figure.}. Much of the data between phases 0.25 and 0.6
were affected by an intermittent telescope control problem 
that grossly degraded the images.
The light curve models give acceptable fits for $i = 50\,\pm\,5^\circ$; the 
curve drawn is for $i = 50$ degrees, $M_1 = 0.8\ {\rm M}_{\odot}$, $M_2 = 0.5\ {\rm M}_{\odot}$, 
and a secondary star with $T_{\rm eff} = 3800$ kelvin and extra light equivalent
to $3 \times 10^{-17}$ erg cm$^{-2}$ s$^{-1}$ \AA $^{-1}$.  In Figure \ref{fig:62cyg_mass}, we show the mass constraints for V1062 Cyg for a variety of potential inclinations.

\section{Discussion} 
\label{sec:disc}

For ASASSN-14ho, we find the orbital period $P = 350.14\,\pm\,0.15$\,min and determine the secondary's spectral type to be K4 $\pm$ 2 subclasses. By modeling the light curve, we estimate $i = 52\,\pm\,5^{\circ}$.  The mass function and inclination suggest white dwarf mass $M_1 > 0.8$ M$_{\odot}$.

For V1062 Cyg, we find $P = 348.25\,\pm\,0.60$\,min, and classify the secondary as M0.5. The light curve yielded $i = 50\,\pm\,5^{\circ}$.

Both systems are dwarf novae with late-type secondary stars that dominate the optical flux. Both systems have orbital periods very close to 5.8 hr, well above the period gap of 2-3\,hrs \citep{Warner95CVs}. The period gap is conventionally understood to represent a transition between two mass accretion stages in the CV's lifetime \citep{Howell01Pgap}.  

Despite their nearly-identical orbital periods, the two systems' secondary stars differ significantly in spectral type. \citet{knigge11} tabulate a semi-empirical sequence of the mean properties of CV secondaries, which gives $M_2 = 0.67 {\rm M_{\odot}}$ and a spectral type of M0 for typical
secondaries at this period.  Our observations of V1062 Cyg are consistent with this, but the
secondary of ASSASN-14ho is somewhat warmer and significantly less massive than their fiducial
sequence.  The empirical donor star data tabulated by \citet{knigge06} does include secondaries as warm as that of ASASSN-14ho near this period.

The H$\alpha$ line profile of V1062 Cyg shows an unusual narrow component that moves in phase with the secondary star, as well as a fainter diffuse component that moves approximately with the white dwarf and accretion disk. This may arise from irradiation of one face of the secondary star.  Very narrow H$\alpha$ lines phased with the secondary are sometimes seen in VY Scl stars during extreme low states (see e.g. \citealt{Weil18} and references therein).  Far-ultraviolet observations might clarify whether the white dwarf or accretion structures
in this system are unusually hot.


\acknowledgments 
\textit{Acknowledgments:} This work uses data collected at the
SAAO 1.0\,m and 1.9\,m and the MDM 1.3\,m and 2.4\,m telescopes. It is
supported in part by the University of Cape Town and Dartmouth College's Frank
J. Guarini Institute for International Education. We would also like to thank
both Kathryn E. Weil and Meredith Joyce for their generous consultations. We
thank the entire Dartmouth 2017 Astronomy Foreign Study Program: Carter H.
Bartram, Abigail J. Buckley, Michael D. Cobb, Ana M. Colon, James G. Detweiler,
John M. Elliot, Emily A. Golitzin, Alana M. Juric, Alexander S. Putter, and
Anne M. Woronecki, for their unwavering support and assistance.
This publication makes use of data products from the AAVSO
Photometric All Sky Survey (APASS), funded by the Robert Martin Ayers
Sciences Fund and the National Science Foundation. The authors H. Breytenbach, 
M. Motsoaledi, and P. Woudt acknowledge support through the National Research 
Foundation of South Africa; Kerry Paterson acknowledges funding by the 
National Research Foundation of South Africa
(NRF) through a South African Radio Astronomy Observatory (SARAO) bursary,
and University of Cape Town (UCT).  We 
would like to thank Erek Alper for taking the photometric MDM
observations of V1062 Cygni.  Finally, we thank the anonymous
referee for their careful reading and suggestions that significantly 
improved this paper.

\bibliographystyle{yahapj}
\bibliography{ref}

\end{document}